# Control of the differential interference contrast in reinjected bimode laser.

ERIC LACOT[1,2,*], OLIVIER JACQUIN[1,2], OLIVIER HUGON[1,2] AND HUGUES GUILLET DE CHATELLUS[1,]

1University 1Univ. Grenoble Alpes, LIPhy, F-38000 Grenoble, France
2CNRS, LIPhy  F-38000 Grenoble,France
*Corresponding author: eric.lacot@ujf-grenoble.fr

**We have demonstrated both theoretically and experimentally that it is possible to control (i.e. to enhance or to cancel) the contrast of the interference pattern appearing in the intensity images obtained with a Laser Optical Feedback Imaging (LOFI) setup using a bimode laser. The laser is composed of two coupled orthogonally polarized states which interact (i.e. interfere) through the cross saturation laser dynamics. The contrast control is made by choosing the frequency-shift (i.e. the beating frequency) between the feedback electric fields and the intracavity electric fields. We show that the interference contrast of the output power modulation of the laser total intensity is independent from the frequency-shift and is always maximal. On the other hand, the interference contrast of each polarization state is frequency dependent. The maximal contrast is obtained when the frequency shift is equal to one of the resonance frequencies of the bimode dynamics, and is very low (and almost cancels) for an intermediate frequency located at the intersection of the two resonance curves.**

## 1. Introduction

With the optical setups based on the detection of ballistic photons (which have not experienced a scattering event), while we are able to obtain high optical resolution ($\approx \mu m$), the accessible depth is quickly limited ($\approx mm$) by the small number of remaining photons. For example, optical setups like Optical Coherence Tomography (OCT) [1,2] or confocal microscopy [3], belong to this family. To improve the in-depth resolution, the challenge is therefore to increase the sensitivity of such methods, but not at the price of expensive equipment and of complex optical alignment. To overcome this problem, one solution is to use the laser optical feedback. Indeed, since the pioneer work of K. Otsuka on self-mixing modulation effects in class-B laser [4], the sensitivity of laser dynamics to frequency-shifted optical feedback has been used in autodyne interferometry and metrology [5], for example in self-mixing laser Doppler velocimetry [6-9], vibrometry [10-12], near field microscopy [13,14] and laser optical feedback imaging (LOFI) experiments [15-20]. Compared to conventional optical heterodyne detection, frequency-shifted optical feedback shows an intensity modulation contrast higher by several orders of magnitude and the maximum of the modulation is obtained when the shift frequency is resonant with the laser relaxation oscillation frequency [21]. In this condition, an optical feedback level as low as -170 dB (i.e. $10^{17}$ times weaker than the intracavity power) has been detected [7].

In previous papers [22-23], we have demonstrated that in LOFI interferometry, the main advantage of the resonant gain (defined by the ratio between the cavity damping rate and the population-inversion damping rate of the laser) is to raise the laser quantum noise over the detector noise in a relatively large frequency range around to the laser relaxation frequency, leading to a shot noise limited signal to noise ratio (SNR). For high-speed imaging, the signal acquisition time must be decreased, leading to a reduction of the SNR of our LOFI setup. To overcome this problem, the laser output power could be increased, but this often leads to a multimode behavior of the laser and more particularly to a coupled dynamical behavior (principally due to spatial hole burning) in a microchip solid-state laser. Due to the multimode behavior of the laser, the LOFI image, obtained from the measurement of the modulation amplitude of the laser intensity, can exhibit interference pattern. More specifically, in our bimode microchip laser which runs on two orthogonal states of polarization, interferences come from the interaction (i.e the superposition) of the intensity modulation of the two laser polarization states, via the cross coupling laser dynamics [24-27].

The main objective of this paper is to show how the cross coupling dynamics of a bimode laser can be used to control (enhance or cancel) the contrast of the interference patterns in LOFI amplitude images. To our knowledge, this paper is also the first demonstration of a LOFI setup working with a bimode laser.

This paper is organized as follows. Firstly, after a basic description of our LOFI setup working with a bimode laser, we give the model describing our laser working with two orthogonal states of polarization coupled trough the cross saturation laser dynamics. In our model, each polarization is submitted to a specific frequency-shifted optical feedback. From this model, the analytical expression of the LOFI signal, (i.e. the amplitude and phase of the beating) is obtained. Then the theoretical contrast of the interference pattern, induced by the cross saturation laser dynamics and due to the phase difference between the two reinjected polarization states (and therefore by the phase difference between the two intensity modulations) is determined for each polarization state and also for the total intensity of the laser. Then, the frequency-shift which allows canceling the interference contrast is determined. Finally, the analytical predictions are confirmed by numerical simulations and by the acquisition of images in different experimental conditions.

## 2. LOFI with a bimode laser

### A. LOFI Setup

A schematic diagram of the LOFI experimental setup is shown in Fig. 1. The laser is a diode pumped Nd:YAG microchip laser. The maximum available pump power is 380 mW at 810 nm, the threshold pump power of a laser is 75mW and the total output power of the microchip laser is 80 mW with a central wavelength of 1064 nm. The laser cavity is a plane-parallel cavity which is stabilized by the thermal lens induced by the Gaussian pump beam. The two dielectric mirrors are directly coated on the laser material (full cavity). The input dichroic mirror allows to transmit the pump power and to totally reflect the infrared laser wavelength. On the other side, the dichroic output mirror allows to totally reflect the pump power (to increase the pump power absorption and therefore the laser efficiency) and only partially reflects (95%) the laser wavelength. The microchip cavity is relatively short $L_c \approx 1 \times 10^{-3} m$. This ensures a high damping rate of the cavity and therefore a good sensitivity to optical feedback. Due to small birefringence induced by a small mechanical pressure on the laser crystal, the optical spectrum of the laser is composed of two wavelengths corresponding to orthogonal polarization states (indexed 0 and 90 in the following) [24-27]. The central wavelength and the wavelength difference are respectively given by $\lambda_m = \frac{\lambda_0 + \lambda_{90}}{2} = 1064 \times 10^{-9} m$ and $\Delta\lambda = \lambda_0 - \lambda_{90} = 0.24 \times 10^{-9} m$. The total output power of the microchip bimode laser is approximately equally divided between the two orthogonal polarization states. This ratio can be controlled by adjusting the orientation of the linear polarization of the pump.

The laser beam is sent on the target through a frequency shifter (acousto-optic device). A part of the light diffracted and/or scattered by the target is then reinjected inside the laser cavity after a second pass through the frequency shifter. Therefore, the optical frequencies of the reinjected light are shifted by $F_e$. This frequency can be adjusted and is typically of the order of the laser relaxation frequencies which are in the megahertz range for our microchip laser.

The optical feedback is characterized by the complex target reflectivity ($r_i = \sqrt{R_i} \exp(j\Phi_i)$ with $i = 0$ or $i = 90$), where the phase $\Phi_i = \frac{2\pi}{\lambda_i} c\tau_e$ describes the optical phase shift induced by the round trip time delay $\tau_e$ (i.e. the distance $d_e = c\tau_e$, where $c$ is the velocity of light) between the laser and the target. The effective power reflectivity ($R_i = |r_i|^2$) takes into account the target albedo, the numerical aperture of the collection optics, the frequency shifters efficiencies, the transmission of all optical components and the overlap of the retro-diffused field with the Gaussian cavity beam (confocal feature).

The coherent interaction (beating) between the lasing electric fields and the frequency-shifted reinjected fields leads to a modulation of the laser output power at $F_e$. For the detection purpose, the laser output beam is split into two beams. The first one is used to record the dynamics of each polarization state through a Wollaston prism, while the second one is used to record the dynamics of the total intensity of the laser. The voltage delivered by the photodiodes are finally analyzed by a numerical oscilloscope which allows Fast Fourier Transform (FFT) calculations, and processed by a lock-in amplifier which gives the LOFI signal (i.e. the amplitude and the phase of the retro-diffused electric field) at the demodulation frequency $F_e$.

Experimentally, the LOFI images are obtained pixel by pixel (i.e. point by point, line after line) by a full 2D galvanometric scanning.

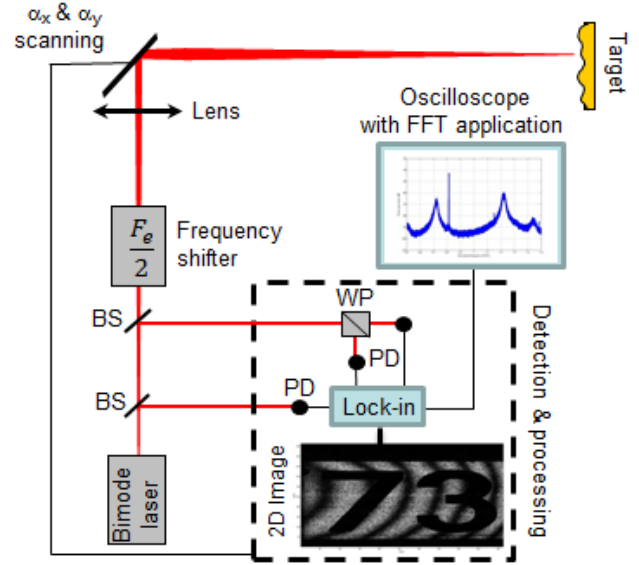

Fig. 1 (Color online) Schematic diagram of the LOFI setup using a bimode laser. BS: Beam Splitter, WP: Wollaston Prism, PDs: Photodiodes. The bimode laser in a Nd:YAG microchip laser with two orthogonal polarization states which are cross-coupled by the saturation of laser gain. The image is obtained pixel by pixel by using a 2D galvanometric scanning.

At this point, one can already notice that compared to a conventional heterodyne interferometer, the LOFI setup shown here does not require complex alignment. Indeed, the LOFI interferometer is even always self-aligned because the laser simultaneously fulfills the functions of the source (i.e. photons-emitter) and of the photo-detector (i.e. photons-receptor).

### B. LOFI Modeling

In the case of a weak frequency shifted optical feedback reinjected into a bimode laser ( $R_i \ll 1$ ) and of a short round trip time delay ( $F_e \tau_e \ll 1$ ), the dynamical behavior of the bimode laser can be described by the following set of differential equations [15,21,26-28]:

$$\frac{dI_0}{dt} = BI_0 N_0 - \gamma_c I_0 + \gamma_c \sqrt{R_0} I_0 \cos(\Omega_e t + \Phi_0) + L_{I_0}(t) \quad \textbf{(1a)}$$

$$\frac{dN_0}{dt} = \gamma_1 \left[ N_p - N_0 \right] - BN_0 \left( I_0 + \alpha I_{90} \right) + L_{N_0}(t) \quad \textbf{(1b)}$$

$$\frac{dI_{90}}{dt} = gBI_{90} N_{90} - \gamma_c I_{90} + L_{I_{90}}(t) + \gamma_c \sqrt{R_{90}} I_{90} \cos(\Omega_e t + \Phi_{90}) \quad \textbf{(1c)}$$

$$\frac{dN_{90}}{dt} = \gamma_1 \left[ N_p - N_{90} \right] - gBN_{90} \left( I_{90} + \alpha I_0 \right) + L_{N_{90}}(t) \quad \textbf{(1d)}$$

where $I_i$ and $N_i$ are respectively the laser intensity (photon unit) and the population inversion of the polarization state $i=0$ and $i=90$, which are coupled through the cross-saturation parameter $\alpha$

In Eqs. (1), each cosine function expresses the coherent interaction (i.e. the beating at the angular frequency: $\Omega_e = 2\pi F_e$ ) between the lasing and the optical feedback electric field. Regarding the noise, the laser quantum fluctuations are described by the Langevin noise functions $L_{N_i}(t)$ and $L_{I_i}(t)$, which have a zero mean value and a white noise type correlation function [29-30].

In Eqs.(1), $\gamma_1$ is the decay rate of the population inversion, $\gamma_c$ is the laser cavity decay rate, $\gamma_1 N_p$ is the pumping rate and $B$ is related to the Einstein coefficient (i.e. the laser cross section), where $g$ describes the cross section ratio of the two polarization states. In the present paper, the case of a bimode laser with a symmetrical gain ( $g=1$ ) will be studied both analytically and numerically, while the case of the asymmetrical gain ( $g \neq 1$ ) will only be processed numerically. This choice has been made in order to avoid heavy (but straightforward) analytical calculations which are not necessary for the physical interpretation of our experimental results. Indeed, the laser that has been used in the experimental section is a bimode laser with a very slightly asymmetrical gain ( $g \approx 1$ ).

For a symmetrical gain ( $g=1$ ), the steady-state of Eqs. (1) is simply given by:

$$N_{0,s} = N_{90,s} = \frac{\gamma_c}{B} = N_s \quad \textbf{(2a)}$$

$$I_{0,s} = I_{90,s} = \frac{1}{1+\alpha} I_{sat} [A-1] = I_s \quad \textbf{(2b)}$$

where $A = \frac{N_p}{N_s}$ is the normalized pumping parameter and $I_{sat} = \frac{\gamma_1}{B}$ is linked to the saturation intensity of the laser transition.

The two orthogonal polarization states of the laser have therefore the same intensity, making it possible to obtain interferences (between the two laser intensity modulations induced by the optical feedback) with a maximum contrast (i.e. a high visibility).

To determine the small modulation of the laser intensity induced by the optical feedback, the set of Eqs.(1) are linearized around the steady state given by Eqs. (2):

$$I_i(t, \Delta\Phi) = I_s + \Delta I_i(t, \Delta\Phi) \quad \textbf{(3a)}$$

$$N_i(t, \Delta\Phi) = N_s + \Delta N_i(t, \Delta\Phi) \quad \textbf{(3b)}$$

with $\Delta I_i(t, \Delta\Phi) \ll I_s$ , $\Delta N_i(t, \Delta\Phi) \ll N_s$ and where $\Delta\Phi = \Phi_0 - \Phi_{90} = 2\pi \frac{\Delta\lambda}{\lambda_m^2} 2 d_e$ is the phase difference between the two feedback modulations induced by the phase difference between the two reinjected electric fields. Then a Fourier transformation allows to convert the differential equations into linear algebraic equations [21,29]. Finally, their resolutions give us the complex modulation amplitudes of the polarization state intensities ( $\Delta \tilde{I}_i(\Omega_e, \Delta\Phi)$ ) at the modulation angular frequency $\Omega_e$:

$$\Delta \tilde{I}_0(\Omega_e, \Delta\Phi) = \frac{\gamma_c I_s \left[ i\Omega_e + \gamma_1 A \right] \exp(i\Phi_{90}) \left[ \sqrt{R_0} \exp(i\Delta\Phi) \left[ \Omega_R^2 - \Omega_e^2 + i\Omega_e \gamma_1 A \right] - \sqrt{R_{90}} \alpha \Omega_R^2 \right]}{\left[ \Omega_{R,+}^2 - \Omega_e^2 + i\Omega_e \gamma_1 A \right] \left[ \Omega_{R,-}^2 - \Omega_e^2 + i\Omega_e \gamma_1 A \right]}, \quad \textbf{(4a)}$$

$$\Delta \tilde{I}_{90}(\Omega_e, \Delta\Phi) = \frac{\gamma_c I_s \left[ i\Omega_e + \gamma_1 A \right] \exp(i\Phi_{90}) \left[ \sqrt{R_{90}} \left[ \Omega_R^2 - \Omega_e^2 + i\Omega_e \gamma_1 A \right] - \sqrt{R_0} \exp(i\Delta\Phi) \alpha \Omega_R^2 \right]}{\left[ \Omega_{R,+}^2 - \Omega_e^2 + i\Omega_e \gamma_1 A \right] \left[ \Omega_{R,-}^2 - \Omega_e^2 + i\Omega_e \gamma_1 A \right]}, \quad \textbf{(4b)}$$

with $\Omega_{R,+}^2 = \gamma_c \gamma_1 (A-1)$, $\Omega_{R,-}^2 = \Omega_{R,+}^2 \frac{(1-\alpha)}{1+\alpha}$ and $\Omega_R^2 = \frac{\Omega_{R,+}^2 + \Omega_{R,-}^2}{2}$ .

At this point, one can notice that the cross coupling parameter can be simply extracted from the knowledge (i.e. from the experimental measurements) of the two angular eigenfrequencies of the laser dynamics. The cross coupling parameter is given by the following relation: $\alpha = \frac{\Omega_{R,+}^2 + \Omega_{R,-}^2}{\Omega_{R,+}^2 + \Omega_{R,-}^2}$

By summing and subtracting Eqs. (4), one obtains:

$$\Delta \tilde{I}_+(\Omega_e, \Delta\Phi) = \Delta \tilde{I}_0(\Omega_e, \Delta\Phi) + \Delta \tilde{I}_{90}(\Omega_e, \Delta\Phi)$$

$$= \Delta \tilde{I}_{tot}(\Omega_e, \Delta\Phi)$$

$$= \frac{\gamma_c I_s [i\Omega_e + \gamma_1 A] \exp(i\Phi_{90})}{\Omega_{R,+}^2 - \Omega_e^2 + i\Omega_e \gamma_1 A}$$

$$\times \frac{\left[\sqrt{R_0} \exp(i\Delta\Phi) + \sqrt{R_{90}}\right]}{\Omega_{R,+}^2 - \Omega_e^2 + i\Omega_e \gamma_1 A}$$

$$= G_+(\Omega_e) \exp(i\Phi_{90})$$

$$\times \left[\sqrt{R_0} \exp(i\Delta\Phi) + \sqrt{R_{90}}\right]$$

(5a)

$$\Delta \tilde{I}_-(\Omega_e, \Delta\Phi) = \Delta \tilde{I}_0(\Omega_e, \Delta\Phi) - \Delta \tilde{I}_{90}(\Omega_e, \Delta\Phi)$$

$$= \frac{\gamma_c I_s [i\Omega_e + \gamma_1 A] \exp(i\Phi_{90})}{\Omega_{R,-}^2 - \Omega_e^2 + i\Omega_e \gamma_1 A}$$

$$\times \frac{\left[\sqrt{R_0} \exp(i\Delta\Phi) - \sqrt{R_{90}}\right]}{\Omega_{R,-}^2 - \Omega_e^2 + i\Omega_e \gamma_1 A}$$

$$= G_-(\Omega_e) \exp(i\Phi_{90})$$

$$\times \left[\sqrt{R_0} \exp(i\Delta\Phi) - \sqrt{R_{90}}\right]$$

(5b)

One can observe that $\Delta \tilde{I}_+(\Omega_e, \Delta\Phi)$ and $\Delta \tilde{I}_-(\Omega_e, \Delta\Phi)$ are the complex amplitudes of the eigenmodes of the coupled laser dynamics, with their respective angular eigenfrequencies (i.e. resonance angular frequencies) given by $\Omega_{R,+}$ and $\Omega_{R,-}$.

More particularly, Eq. (5a) shows that the modulation of the total laser intensity is therefore an eigenmode of the laser dynamics. At this point, one can also notice that, in our particular case (g≈1), the intermediate angular frequency previously defined by $\Omega_R^2 = \frac{\Omega_{R,+}^2 + \Omega_{R,-}^2}{2}$, corresponds to the angular frequency for which the two resonance curves, defined by the eigenmodes, intersect each other and therefore have the same modulus: $|G_-(\Omega_R)| = |G_+(\Omega_R)|$.

### C. Theoretical interference contrast

When $\Delta\Phi$ is scanned over two pi, one can determine the interference contrast due to the interaction (i.e. the superposition) of the intensity modulation of the two laser polarization states, via the cross-coupling laser dynamics. More precisely, from the power spectra $|\Delta \tilde{I}_i(\Omega_e, \Delta\Phi)|^2$, the interference contrast is determined by using:

$$C_i(\Omega_e) = \frac{|\Delta \tilde{I}_i(\Omega_e, \Delta\Phi)|^2_{max} - |\Delta \tilde{I}_i(\Omega_e, \Delta\Phi)|^2_{min}}{|\Delta \tilde{I}_i(\Omega_e, \Delta\Phi)|^2_{max} + |\Delta \tilde{I}_i(\Omega_e, \Delta\Phi)|^2_{min}} \quad (6)$$

with $i=0$, $i=90$ for the orthogonal polarization states and $i=tot$ for the total intensity.

From Eqs. (4) and (5), one easily obtains:

$$C_{0(90)}(\Omega_e) = 2 \frac{\sqrt{R_{0(90)}\left(\left(\Omega_R^2 - \Omega_e^2\right)^2 + \left(\Omega_e \gamma_1 A\right)^2\right)} \sqrt{R_{90(0)}\left(\alpha \Omega_R^2\right)^2}}{R_{0(90)}\left[\left(\Omega_R^2 - \Omega_e^2\right)^2 + \left(\Omega_e \gamma_1 A\right)^2\right] + R_{90(0)}\left(\alpha \Omega_R^2\right)^2} \quad (7a)$$

$$C_{tot} = \frac{2\sqrt{R_0 R_{90}}}{R_0 + R_{90}} \quad (7b)$$

Eqs. (7), show that the interference contrast of the total intensity is independent of the frequency-shift ($\Omega_e$), while the interference contrasts of both polarization states are frequency dependent.

At this point, one can notice that when we detect only one polarization state, the interference interaction between the two feedback modulations comes from the coupling by the laser dynamics ($\alpha \neq 0$). Indeed Eq. (7a) gives $C_{0(90)}(\Omega_e) = 0$ when $\alpha = 0$.

To study the dynamical effects of the laser dynamics on the interference contrast, we focus the discussion on the particular case where the amount of optical feedback in each polarization state is approximatively the same ($R_0 \approx R_{90}$) and therefore has no impact on the reduction of the interference contrast (i.e. the fringes visibility).

Under this condition and for a class-B laser $(\Omega_R >> \gamma_1 A)$, Eqs. (7) give:

$$C_{0(90)}(\Omega_e << \Omega_R) \approx 2\left(\frac{\alpha}{1+\alpha^2}\right) \approx 0.7 \quad (8a)$$

$$C_{0(90)}(\Omega_{R,-}) \approx 1 \quad (8b)$$

$$C_{0(90)}(\Omega_R) \approx 2\left(\frac{\gamma_1 A}{\Omega_R \alpha}\right) \approx 10^{-3} << 1 \quad (8c)$$

$$C_{0(90)}(\Omega_{R,+}) \approx 1 \quad (8d)$$

$$C_{0(90)}\left(\Omega_e \gg \Omega_R\right) \approx 2\left(\frac{\alpha \Omega_R^2}{\Omega_e^2}\right) \ll 1 \quad \textbf{(8e)}$$

$$C_{tot} \approx 1 \quad \textbf{(8f)}$$

For both polarization states, Eqs. (8) show that the interference contrast is high at low frequency and is maximal if we detect the feedback interaction (i.e. the beating) at the resonance angular frequencies ($\Omega_{R,-}$ or $\Omega_{R,+}$). On the other hand, the interference contrast is very low (and almost cancels) for the intermediate angular frequency $\Omega_R$ [Eq. (8c)], or when working far away from the resonance zone [Eq. (8e)]. For the total intensity the interference contrast is always maximal. One can notice that all these results can be interpreted by a 180 degrees phase shift across the resonance eigenfrequencies.

Fig. 2 shows the normalized amplitude of the laser output power modulation versus the frequency shift ($F_e = \Omega_e/2\pi$) and the interference condition ($\Delta\Phi$) induced by the target distance.

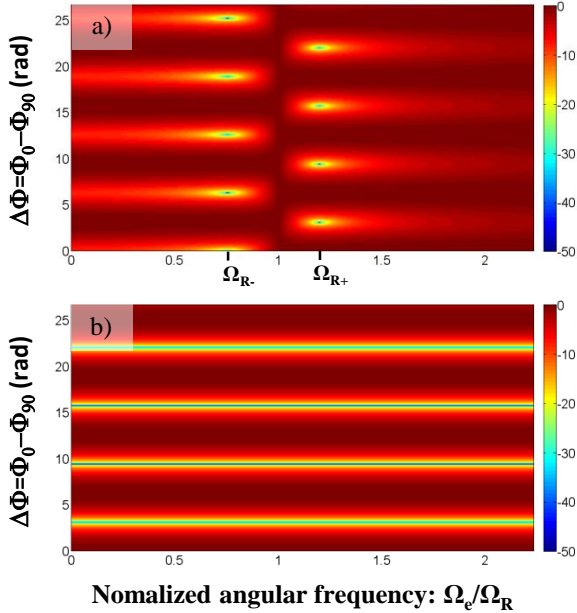

**Nomalized angular frequency: $\Omega_e/\Omega_R$**

Fig. 2. (Color online) Normalized amplitude of the laser output power modulation (in a logarithmic false color scale) versus the normalized angular frequency: $\Omega_e/\Omega_R$ and the phase difference $\Delta\Phi$. For each angular frequency: ($\Omega_e$), the amplitude of the modulation is normalized to the brightest interference condition according to $\Delta\Phi$.

a) Laser horizontal polarization: $\left[\left|\Delta\tilde{I}_0(\Delta\Phi)\right|/\max\left(\left|\Delta\tilde{I}_0(\Delta\Phi)\right|\right)\right]_{\Omega_e}$,

b) Laser total intensity: $\left[\left|\Delta\tilde{I}_{tot}(\Delta\Phi)\right|/\max\left(\left|\Delta\tilde{I}_{tot}(\Delta\Phi)\right|\right)\right]_{\Omega_e}$. Laser parameters: $\gamma_1 = 2.6\times 10^4\ s^{-1}$, $\gamma_c = 2.6\times 10^9\ s^{-1}$, $A = 2.5$, $\alpha = 0.43$. $\Omega_{R,-}/\Omega_R = \sqrt{1-\alpha} = 0.75$, $\Omega_{R,+}/\Omega_R = \sqrt{1+\alpha} = 1.20$. Target parameters: $R_0 = R_{90}$.

In agreement with Eq. (8f), one can observe on Fig. 2(b) that the interference fringe pattern of the total intensity is independent of the frequency shift and that the visibility of the fringes (i.e. the contrast) is always the same. When detecting the polarization states separately, Fig. 2(a) shows that the interference fringe contrasts are maximal at $\Omega_e = \Omega_{R,-}$ and $\Omega_e = \Omega_{R,+}$, while they vanishes (i.e. have no dependence with the target distance) when working at the intermediate angular frequency $\Omega_e = \Omega_R$ or when working far away from the resonance zone $\Omega_e \gg \Omega_R$. At this point one can also notice the phase shift of $\pi$ between the interference pattern at $\Omega_{R,-}$ and $\Omega_{R,+}$. Indeed, the bright fringes at $\Omega_{R,-}$ correspond to the dark fringes at $\Omega_{R,+}$ and vice versa. Between the two possibilities which allow canceling the interference effect between the intensity modulations of the two polarization states, the first solution [Eq. (8c)] is the best. Indeed, the LOFI SNR is shot noise limited only in the vicinity of the resonance frequencies of the laser [22, 23].

Working at this particular angular frequency: ($\Omega_e = \Omega_R$), and coming back to the more general situation where $R_0$ is not necessary equal to $R_{90}$, Eqs. (4) and (5) give:

$$\left|\Delta\tilde{I}_0(\Omega_R, \Delta\Phi)\right|_{\Omega_R \gg \gamma_1 A} \approx \frac{\gamma_c I_s}{\alpha \Omega_R}\sqrt{R_{90}} \quad \textbf{(9a)}$$

$$\left|\Delta\tilde{I}_{90}(\Omega_R, \Delta\Phi)\right|_{\Omega_R \gg \gamma_1 A} \approx \frac{\gamma_c I_s}{\alpha \Omega_R}\sqrt{R_0} \quad \textbf{(9b)}$$

$$\left|\Delta\tilde{I}_{tot}(\Omega_R, 0)\right|_{\Omega_R \gg \gamma_1 A} \approx \frac{\gamma_c I_s}{\alpha \Omega_R}\left[\sqrt{R_0} + \sqrt{R_{90}}\right]$$
$$= \left|\Delta\tilde{I}_0(\Omega_R, 0)\right| + \left|\Delta\tilde{I}_{90}(\Omega_R, 0)\right| \quad \textbf{(9c)}$$

$$\left|\Delta\tilde{I}_{tot}(\Omega_R, \pi)\right|_{\Omega_R \gg \gamma_1 A} \approx \frac{\gamma_c I_s}{\alpha \Omega_R}\left[\left|\sqrt{R_0} - \sqrt{R_{90}}\right|\right]_{R_0 \approx R_{90}} \approx 0 \quad \textbf{(9d)}$$

In agreement with Eq.(8c) (i.e. $C_{0(90)}(\Omega_R) \approx 0$), Eqs. (9a) and (9b) show that the amplitude of the modulation of both polarization states is independent of the phase difference $\Delta\Phi$. At the intermediate frequency, the LOFI signal for each polarization state is therefore insensitive to the interference condition between the two polarization states. More surprisingly (because less intuitive), Eq. (9a), shows that the LOFI signal of the horizontal polarization ($\left|\Delta\tilde{I}_0\right|$) is only induced by the feedback in the vertical polarization ($R_{90} \neq 0$) and vice versa for Eq. (9b).

On the other hand and in agreement with Eq. (8a) ($C_{tot}(\Omega_R) \approx 1$), Eqs. (9c) and (9d) show that the value of the LOFI signal of the total intensity is strongly dependent on the phase difference $\Delta\Phi$ and can disappear for destructive interference condition [see Eq. (9d)]. One can also observe that for constructive interference ($\Delta\Phi = 0$), the amplitude of the LOFI signal of the total intensity is the sum of the LOFI signal of the two orthogonal states [see Eq. (9c)]. Therefore, by working at the intermediate frequency and by detecting both modes separately instead of directly detecting the total intensity, one can obtain a LOFI signal of the same value (no loss of energy), but without the interference fringe pattern (i.e. without the dependence of $\Delta\Phi$).

## 3. Cancellation of the interference fringes

### A. Numerical simulations

Fig. 3 shows numerical results obtained by using a RungeKutta method to solve (without any approximation) the differential equations (1). This numerical simulation is made with a frequency shift equal to the intermediate frequency ($F_e = \frac{\Omega_e}{2\pi} = \frac{\Omega_R}{2\pi} = F_R$) and an anisotropic target ($R_0 \neq 0$ and $R_{90} = 0$). To describe realistic experimental conditions, the laser quantum noise is taken into account in this numerical simulation by programming the Langevin forces.

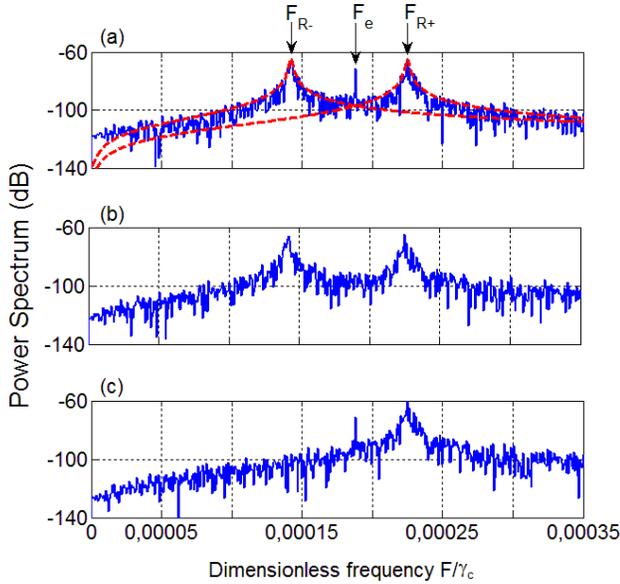

Fig. 3 (Color online) Numerical simulation: square modulus of the Fast Fourier Transform (FFT) of the temporal evolution the bimode laser dynamics, versus the dimensionless frequency $\frac{F}{\gamma_c}$.

a) $\left|FFT\left[\Delta I_{90}(t,\Delta\Phi)/I_{sat}\right]\right|^2$; b) $\left|FFT\left[\Delta I_0(t,\Delta\Phi)/I_{sat}\right]\right|^2$;

c) $\left|FFT\left[\Delta I_{tot}(t,\Delta\Phi)/I_{sat}\right]\right|^2$. Laser parameters: $A = 1.2$, $\gamma_1/\gamma_c = 10^{-5}$, $g = 1$, $\alpha = 0.43$, $F_{R-}/\gamma_c = 1.421 \times 10^{-4}$, $F_{R+}/\gamma_c = 2.251 \times 10^{-4}$, $F_e/\gamma_c = \sqrt{\frac{(F_{R+}^2 + F_{R-}^2)}{2}}/\gamma_c = 1.882 \times 10^{-4}$.
Target parameters: $R_0 \neq 0$ and $R_{90} = 0$. $\Delta\Phi$: any values.

For the LOFI signal, one can observe that the beating frequency appears on the total intensity [see Fig. 3(c)] and also on the vertical polarization state [see Fig. 3(a)] despite the fact that there is no feedback on this state ($R_{90} = 0$). In agreement with Eqs. (9b) and (9c), one can also observe that the two peaks have the same amplitude. On the other hand, one can observe on Fig. 3(b) that no modulation appears at the beating frequency in the horizontal state ($i = 0$) despite the fact that feedback is only on this mode ($R_0 \neq 0$). As already mentioned, this counter-intuitive result is induced by the coupling between the dynamical behaviors of the two polarization states. This result is in agreement with Eq. (9a). In our numerical simulation, the Langevin forces allow to stimulate the noise of the laser dynamics at all frequencies and one can observe that the two polarization states resonate for both frequencies $F_{R,-} = \frac{\Omega_{R,-}}{2\pi}$ and $F_{R,+} = \frac{\Omega_{R,+}}{2\pi}$ [see Figs. 3(a) and 3(b)] while the total intensity, which is an eigenmode of our bimode system, only resonate at the frequency $F_{R,+}$ [see Fig. 3(c)].

On Fig. 3(a), one can also observe that the intermediate frequency which is a very particular frequency for the laser dynamical behavior corresponds to the frequency where the two resonance curves intersect each other.

Fig. 4 shows the variation of the laser output power modulation when the phase difference between the two modes ($\Delta\Phi$) is changed over $4\pi$. This numerical simulation is made with a frequency shift once again equals to the intermediate frequency ($F_e = F_R = \sqrt{\frac{F_{R,+}^2 + F_{R,-}^2}{2}}$) and this time with an isotropic target ($R_0 = R_{90}$). The laser quantum noise is also taken into account in this numerical simulation.

In Fig 4(a), the numerical simulation is made for a bimode laser with a symmetrical gain ($g = 1$). The numerical results can therefore be directly compared with the analytical results given previously. Indeed, in agreement with Eqs (9), Fig. 4(a) shows that the modulation amplitude of both polarization state is independent from the phase difference $\Delta\Phi$. At the intermediate frequency, the LOFI signal for each polarization is therefore insensitive to the interference interaction between them. On the other hand, the LOFI signal of the total intensity is strongly dependent on the phase difference $\Delta\Phi$. One can, also observe on this figure that the sum of the LOFI signal of the two polarizations corresponds the maximum value of the LOFI signal of the total intensity (no energy loss).

In Fig. 4(b), the numerical simulation is made for a bimode laser with an asymmetrical gain ($g \neq 1$). This condition is closer to real experimental situation. Due to the asymmetry, Fig. 4(b) shows a little modulation of the amplitude of both polarization states. Nevertheless, one can observe that the corresponding contrast remains very low as compared to the contrast of the total intensity fringes. In the Fig. 4(b), one can also observe that the interference fringes of the two polarizations have a phase shift of $\pi$. Indeed, the bright fringes of $\left|\Delta\tilde{I}_0\right|$ correspond to the dark fringes of $\left|\Delta\tilde{I}_{90}\right|$ and vice versa. As a consequence, the sum of the LOFI signal of the two polarizations is nearly constant and therefore is independent from the phase difference $\Delta\Phi$ (i.e. independent from the interference condition). The corresponding value once again equals the maximum value (bright fringes) of the LOFI signal of the total intensity [31].

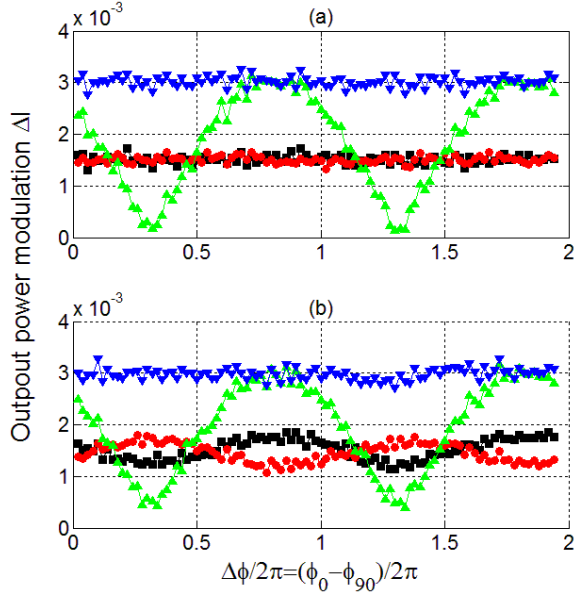

Fig. 4 (Color online) Numerical simulation: amplitude of the laser output power modulation, at the intermediate frequency $F_e = F_R$, versus the phase difference between the two polarization states. Laser parameters: $A = 2.5$, $\gamma_1/\gamma_c = 10^{-5}$, $\alpha = 0.43$. Target parameters: $R_0 = R_{90} = 10^{-12}$. Up triangles: $\left|\Delta\tilde{I}_{tot}(F_e = F_R, \Delta\Phi)\right| = \left|\Delta\tilde{I}_0(F_e = F_R, \Delta\Phi) + \Delta\tilde{I}_{90}(F_e = F_R, \Delta\Phi)\right|$, Squares: $\left|\Delta\tilde{I}_0(F_e = F_R, \Delta\Phi)\right|$, circles: $\left|\Delta\tilde{I}_{90}(F_e = F_R, \Delta\Phi)\right|$, down triangles: $\left|\Delta\tilde{I}_0(F_e = F_R, \Delta\Phi)\right| + \left|\Delta\tilde{I}_{90}(F_e = F_R, \Delta\Phi)\right|$. a) Bimode laser with a symmetrical gain ($g = 1$): $I_{S,0}/I_{sat} = I_{S,90}/I_{sat} = 1.049$, $F_{R+}/\gamma_c = 6.16 \times 10^{-4}$, $F_{R-}/\gamma_c = 3.89 \times 10^{-4}$, $F_R/\gamma_c = 5.15 \times 10^{-4}$. b) Bimode laser with an asymmetrical gain ($g = 0.95$): $I_{S,0}/I_{sat} = 1.077$, $I_{S,90}/I_{sat} = 0.984$, $F_{R+}/\gamma_c = 6.04 \times 10^{-4}$, $F_{R-}/\gamma_c = 3.79 \times 10^{-4}$, $F_R/\gamma_c = 5.04 \times 10^{-4}$.

### B. Experimental LOFI signals

Fig. 5 shows the experimental feedback signal measured by detecting the laser horizontal polarization dynamics. The left column shows the RF power spectra for different experimental conditions. One can principally observe the modulation frequency ($F_e$) induced by the frequency shifted feedback, and also the two resonance frequencies ($F_{R,-}$ and $F_{R,+}$) stimulated by the laser quantum noise. For different values of $F_e$, the right column shows the variation of the LOFI signal (i.e. the amplitude of the modulation of the laser intensity at the modulation frequency $F_e$) when the detected target is moved over a few millimeters on either side of a mean distance of $\langle d_e \rangle \approx 2.5m$.

Firstly, one can observe on the four graphs of the right column, a global bell shape due the confocal filtering of the LOFI microscope, induced by the overlap of the retro-diffused field with the Gaussian cavity beam. Above the global shape, one can also observe that the interference contrast depends on the frequency shift (i.e. modulation frequency).

In agreement with Fig. 2(a), one can observe that the interference contrast is high in the vicinity of the resonance frequencies [Figs 5(a) and 5(c)] and is minimum (and roughly equal to zero) when working at the intermediate frequency [Fig. 5(b)] or far away from the resonance zone [Fig. 5(d)]. In agreement with Fig. 2(a), one can also observe on the right column that the fringes are shifted when we detect the interference fringes on either side of the intermediate frequency $F_R$. Indeed, the minima of Fig. 5(a) and 5(c) are not obtained for the same position of the target.

Fig. 6 shows LOFI images of a slightly tilted target located at a mean distance of $\langle d_e \rangle = 2.5m$ away from the laser. The target is a small part of a car registration plate. The interference fringes are induced both by a small tilt of the registration plate (of the order of 5°) and by the spherical scanning of the laser beam induced by the galvanometric mirrors (see Fig. 1).

The left column shows that the interference fringe pattern is independent from the frequency shift (i.e. of the beating frequency) when we detect the amplitude of the modulation of the total intensity of the bimode laser. On the other hand, when we detect only one mode (the horizontal one in this case), one can observe on the right column that the interference contrast depends on the frequency shift. In agreement with the results of Fig. 5, one can observe that the interference contrast is high in the vicinity of the resonance frequencies [Figs. 6(a) and 6(c)] and is minimum (and roughly equal to zero) when we work at the intermediate frequency [Fig. 6(b)] or far away from the resonance zone [Fig. 6(d)]. In agreement with Fig. 2(a), one can also observe on the right column that the fringes are shifted when we look at the interference pattern detected on either side of the intermediate frequency $F_R$. Indeed, the bright fringes of Fig. 6(a) correspond to the dark fringes of Fig. 6(c) and vice versa.

Therefore by choosing the frequency shift (i.e. the beating frequency) between the feedback electric field and the intracavity electric field, we are able to control and also to cancel the interference pattern appearing in the LOFI images. This control is induced by the cross coupling of the laser dynamics between the two orthogonal polarizations of our bimode laser.

## 4. Conclusion

We have demonstrated both theoretically and experimentally that it is possible to control (i.e. to enhance or to cancel) the contrast of the interference pattern appearing in the intensity images obtained with a Laser Optical Feedback Imaging (LOFI) setup using a bimode laser.

Firstly, after a basic description of our LOFI setup working with a bimode laser, we have given the model describing our laser running on two coupled polarization states and submitted to frequency-shifted optical feedbacks. From this model, the analytical expression of the LOFI signal is obtained in the case of a bimode laser with a symmetrical gain ($g = 1$). Then, the theoretical contrast of the interference pattern (induced by a phase difference $\Delta\Phi$ between the two reinjected polarizations) is determined for each polarization state and also for the total intensity of the laser. These results show that the contrast can be controlled by choosing the frequency-shift (i.e. the beating frequency) between the feedback electric field and the intracavity electric field. This control is possible due to the cross coupling dynamics between the two polarizations (i.e. $\alpha \neq 0$). We show that the interference contrast of the output power modulation of the laser total intensity is independent from the frequency-shift and is always maximal. On the other hand, the interference contrast for each polarization is frequency dependent. The maximal contrast is obtained when the frequency shift is equal to the resonance frequencies of the bimode dynamics ($F_{R,-} = \Omega_{R,-}/2\pi$ and $F_{R,+} = \Omega_{R,+}/2\pi$), and is very low (and almost cancels) for an intermediate frequency located at the intersection of the two resonance curves $F_R = \sqrt{\dfrac{F_{R,+}^2 + F_{R,-}^2}{2}}$. Another possibility to cancel the interference contrast is to use a beating frequency very far away from the resonance range of the laser dynamics. But this last solution gives a lower SNR for the LOFI images. Then, the theoretical predictions are confirmed by a numerical simulation of the laser dynamics. These simulations show that each polarization state exhibits two resonance frequencies while the total intensity, which is an eingenmode of the laser dynamics, exhibits only the highest relaxation frequency. The numerical simulation also confirm that by working at the intermediate frequency and by detecting both polarizations separately, one can obtain a LOFI signal of the same value (no loss of energy) than the total intensity, but without the interference fringe pattern (i.e. without the dependence with $\Delta\Phi$).

Finally all the predictions are confirmed by the acquisitions of LOFI images where the contrast of the interference fringe pattern can be controlled (enhanced or cancelled) by adjusting the value of the frequency shift and by choosing the detected polarization states. From this work, one can suppose that the same mode-dependent modulation property is expected in general multimode solid-state lasers besides dual-polarization lasers [32].

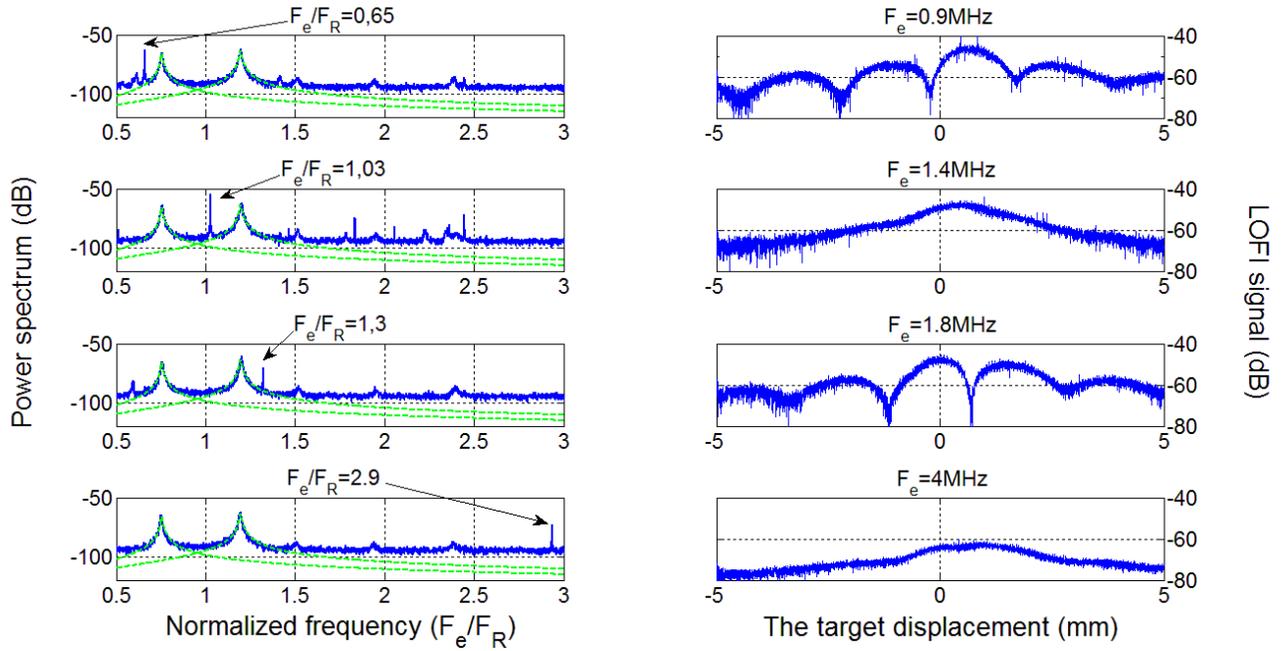

Fig. 5: Experimental feedback signals for a target located at a mean distance of $\langle d_e \rangle = 2.5m$ away from the laser. Left column: power spectrum of the laser horizontal polarization dynamics: $\left| FFT\left[ I_0\left(t, \Delta\Phi(\langle d_e \rangle)\right)\right]\right|^2$ Right column: output power modulation, at the demodulation frequency $F_e$ (i.e. LOFI Signal), of the laser horizontal polarization: $\left|\Delta\tilde{I}_0\left(F_e, \Delta\Phi(d_e)\right)\right|$, versus the relative target displacement: $\Delta d_e = d_e - \langle d_e \rangle$. The global shape of the signal is due the confocal filtering of the LOFI microscope induced by the overlap of the retro-diffused field with the gaussian cavity beam. Frequency-shift: (a) $F_e = 0.9MHz$ ($F_{R,-} = 1.0MHz$); (b) $F_e = 1.4MHz \approx F_R$; (c) $F_e = 1.8MHz$ ($F_{R,+} = 1.7MHz$); (d) $F_e = 4MHz > F_{R,+}$.

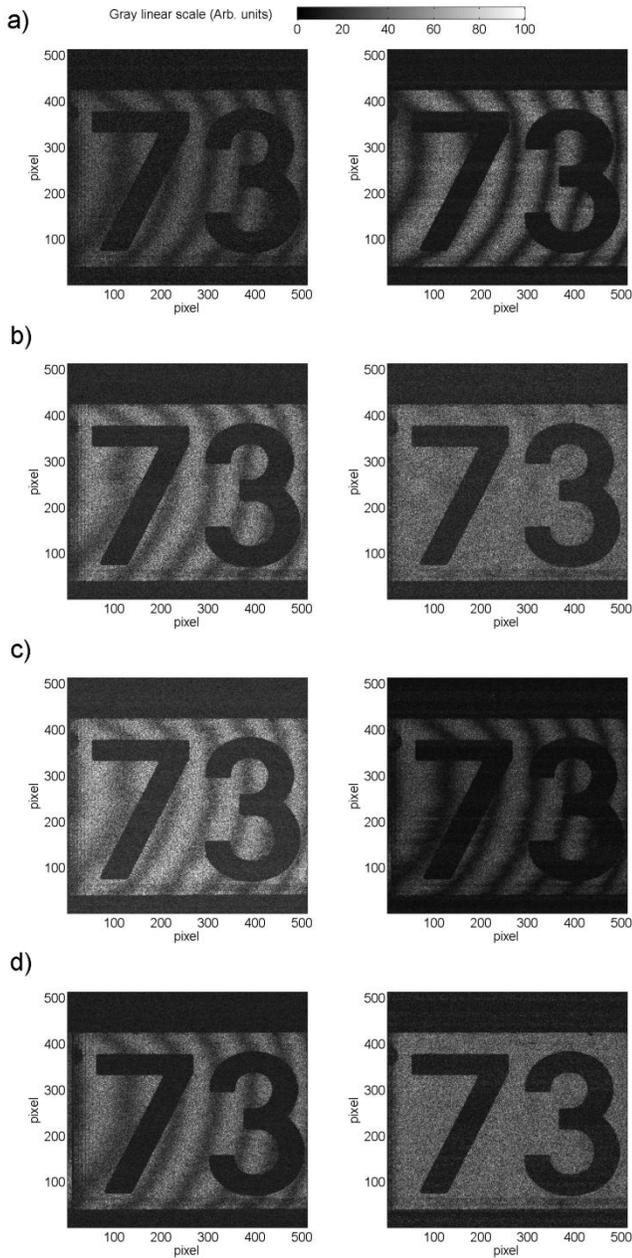

Fig. 6: LOFI images of a target located at a mean distance of $\langle d_e \rangle = 2.5 m$ away from the laser. The target is a small part of a car registration plate. The real size of each 512x512 pixels image is 10 cm x 10 cm. Left column: output power modulation of the laser total intensity: $\left| \Delta \tilde{I}_{tot}(F_e, x, y) \right|$. Right column: output power modulation of the laser horizontal polarization: $\left| \Delta \tilde{I}_0(F_e, x, y) \right|$. In both cases the two wavelengths (i.e. the two polarizations) are send simultaneously on the target. Frequency-shift: (a) $F_e = 0.9 MHz$ ($F_{R,-} = 1.0 MHz$); (b) $F_e = 1.4 MHz \approx F_R$; (c) $F_e = 1.8 MHz$ ($F_{R,+} = 1.7 MHz$); (d) $F_e = 4 MHz > F_{R,+}$.